\documentclass[a4paper,twoside,twocolumn,10pt,titlepage,final]{article}
\usepackage[english]{babel}
\usepackage[figuresright]{rotating}
\usepackage{epsfig}
\usepackage[intlimits,tbtags]{amsmath}
\usepackage{amssymb,amscd}
\usepackage{array}
\usepackage{multirow}
\usepackage{supertabular}
\usepackage{dcolumn}
\usepackage{xspace}
\usepackage{tabularx}
\usepackage{color}
\usepackage{calc}
\usepackage[space,noadjust]{cite}
\chardef\usc=95
\chardef\til=126
\catcode`\@=11 % @ signs are now treated as letters
\DeclareRobustCommand\xdotspace{\futurelet\@let@token\@xdotspace}
\def\@xdotspace{%
  \ifx\@let@token.\else
  \ifx\@let@token\bgroup.\else
  \ifx\@let@token\egroup.\else
  \ifx\@let@token\/.\else
  \ifx\@let@token\ .\else
  \ifx\@let@token~.\else
  \ifx\@let@token!.\else
  \ifx\@let@token,.\else
  \ifx\@let@token:.\else
  \ifx\@let@token;.\else
  \ifx\@let@token?.\else
  \ifx\@let@token/.\else
  \ifx\@let@token'.\else
  \ifx\@let@token).\else
  \ifx\@let@token-.\else
  \ifx\@let@token\@xobeysp.\else
  \ifx\@let@token\space.\else
  \ifx\@let@token\@sptoken.\else
   .\space
   \fi\fi\fi\fi\fi\fi\fi\fi\fi\fi\fi\fi\fi\fi\fi\fi\fi\fi}
\catcode`\@=12 % @ signs are no longer letters

\newcommand{\stru}[2]{%
   \relax\ifmmode\hbox{\vrule height#1 depth#2 width0pt}%
   \else\vrule height#1 depth#2 width0pt\fi}
\newcommand{\Ronum}[1]{\uppercase\expandafter{\romannumeral#1}}
\newcommand{\ronum}[1]{\expandafter{\romannumeral#1}}
\DeclareRobustCommand{\LaTeXZ}{%
  \LaTeX\kern-.05em4\kern-.1em
  {\raisebox{-0.2ex}{$\scriptstyle\text{ZEUS}$}}\xspace}
\DeclareMathAlphabet{\mathbf}{OT1}{cmr}{bx}{sl}
\newcommand{\eVdist}{\kern-0.06667em}

\newcommand{\gev}{{\,\text{Ge}\eVdist\text{V\/}}}

\newcommand{\slashfrac}[2]{%
  \raisebox{0.5ex}{\ensuremath #1}\kern-0.12em/\kern-0.08em
  \raisebox{-.8ex}{\ensuremath #2}}
\newcommand{\sqr}[3]{%
    {\vcenter{\hrule height.#3ex\hbox{\vrule width.#2ex height#1ex
     \kern#1ex\vrule width.#3ex}\hrule height.#2ex}}}

\catcode`\@=11 % @ signs are now treated as letters
\newcommand{\parenbar}{\mathpalette\p@renb@r}
\def\p@renb@r#1#2{\vbox{%
  \ifx#1\scriptscriptstyle \dimen@.7em\dimen@ii.2em\else
  \ifx#1\scriptstyle \dimen@.8em\dimen@ii.25em\else
  \dimen@1em\dimen@ii.4em\fi\fi \offinterlineskip
  \ialign{\hfill##\hfill\cr
    \vbox{\hrule width\dimen@ii}\cr
    \noalign{\vskip-.3ex}%
    \hbox to\dimen@{$\mathchar300\hfil\mathchar301$}\cr
    \noalign{\vskip-.3ex}%
    $#1#2$\cr}}}
\catcode`\@=12 % @ signs are no longer letters

\newcommand{\IP}{{\rm I$\kern-0.01667em$P}\xspace}

\mathchardef\qsm=63
\mathchardef\pls=43
\mathchardef\mns=512
\mathchardef\plm=518
\mathchardef\eql=61
\mathchardef\smallleft=300
\mathchardef\smallright=301
\mathchardef\les=316
\mathchardef\gre=318
\mathchardef\leq=532
\mathchardef\grq=533
\catcode`\@=11 % @ signs are now treated as letters
\newcounter{pict@width}
\newcounter{pict@height}
\newlength{\pict@scale}
\newcommand{\psfigadd}[4]{%
\setcounter{pict@width}{1*\ratio{#2+\pict@scale/2}{\pict@scale}}
\setcounter{pict@height}{1*\ratio{#3+\pict@scale/2}{\pict@scale}}
\setlength{\unitlength}{\pict@scale}
\hbox to #2{\hspace{-\fill}\begin{picture}(\thepict@width,\thepict@height)
\put(0,0){\psfig{figure=#1,width=#2,height=#3,clip=}}
\SetScale{0.283466457}
\SetWidth{1.763889}
{#4}
\end{picture}}
}
\newcounter{pict@widthfst}
\newcounter{pict@widthscd}
\newcounter{pict@widthtot}
\newcommand{\psfigaddtwo}[7]{%
\setcounter{pict@widthfst}{1*\ratio{#2+\pict@scale/2}{\pict@scale}}
\setcounter{pict@widthscd}{1*\ratio{#2+#4+\pict@scale/2}{\pict@scale}}
\setcounter{pict@widthtot}{1*\ratio{#2+#4+#6+\pict@scale/2}{\pict@scale}}
\setcounter{pict@height}{1*\ratio{#3+\pict@scale/2}{\pict@scale}}
\setlength{\unitlength}{\pict@scale}
\hbox{\hspace{-\fill}\begin{picture}(\thepict@widthtot,\thepict@height)
\put(0,0){\psfig{figure=#1,width=#2,height=#3,clip=}}
\put(\thepict@widthscd,0){\psfig{figure=#5,width=#6,height=#3,clip=}}
\SetScale{0.283466457}
\SetWidth{1.763889}
{#7}
\end{picture}}
}
\newcommand{\psfigror}[4]{%
\setcounter{pict@width}{1*\ratio{#2+\pict@scale/2}{\pict@scale}}
\setcounter{pict@height}{1*\ratio{#3+\pict@scale/2}{\pict@scale}}
\setlength{\unitlength}{\pict@scale}
\hbox{\begin{picture}(\thepict@width,\thepict@height)
\put(0,\thepict@height){\psfig{figure=#1,width=#3,height=#2,clip=,angle=270}}
\SetScale{0.283466457}
\SetWidth{1.763889}
{#4}
\end{picture}}
}
\newcommand{\psfigrol}[4]{%
\setcounter{pict@width}{1*\ratio{#2+\pict@scale/2}{\pict@scale}}
\setcounter{pict@height}{1*\ratio{#3+\pict@scale/2}{\pict@scale}}
\setlength{\unitlength}{\pict@scale}
\hbox{\begin{picture}(\thepict@width,\thepict@height)
\put(0,0){\psfig{figure=#1,width=#3,height=#2,clip=,angle=90}}
\SetScale{0.283466457}
\SetWidth{1.763889}
{#4}
\end{picture}}
}
\catcode`\@=12 % @ signs are no longer letters
\newlength\listtextwidth

\catcode`\@=11 % @ signs are now treated as letters
\newlength{\@tabfninsert}
\newlength{\@tabfnwidth}
\newcommand{\tabfootnote}[2]{%
  \setlength{\@tabfninsert}{0.8em}
  \setlength{\@tabfnwidth}{\textwidth}
  \addtolength{\@tabfnwidth}{-\@tabfninsert}
  \addtolength{\@tabfnwidth}{-0.4em}
  \noindent\makebox[\@tabfninsert][r]{\footnotesize$^{#1}$\hfil}\hfill%
  \parbox[t]{\@tabfnwidth}{\footnotesize #2\hfill}}
\catcode`\@=12 % @ signs are no longer letters

\newcommand{\fxp}{FMNR$\otimes$P{\sc ythia }}
\newcommand{\fxpnospace}{FMNR$\otimes$P{\sc ythia}}

\begin{document}
%------------------------------------------------------------------------------
%       Title sheet
%------------------------------------------------------------------------------
%\include{FxP-paper-tit}
%------------------------------------------------------------------------------
%       Text
%------------------------------------------------------------------------------
\pagenumbering{arabic} 
\pagestyle{plain}

% ----------------------------------------------------------------------------
%       Header
% ----------------------------------------------------------------------------
\twocolumn[ 
\noindent {\bf {\LARGE \textsf{The FMNR$\mathbf{\otimes}$P{\sc ythia} interface for Heavy Quark production at HERA}} }\\
\vspace*{0.7cm}

\noindent A.~Geiser$^1$ and A.~E.~Nuncio~Quiroz$^{2,3}$\\

\noindent $^1$ Deutsches Elektronen-Synchroton, DESY, Hamburg, Germany \\
$^2$ Universit\"at Hamburg, Hamburg, Germany\\
$^3$ now at Physikalishes Institut der Universit\"at Bonn, Bonn, Germany \\ 
\vspace{0.5cm}

\noindent{\bf Abstract:}
A method to calculate heavy flavor visible-level cross sections at Next-to-Leading Order (NLO) in Quantum Chromodynamics (QCD), based on an interface of the FMNR program to P{\sc ythia}, is described.
It uses the NLO prediction at quark level provided by FMNR, with a statistical reduction procedure (REDSTAT) that allows a link to P{\sc ythia} 6.2 to be made, from where the description of the full hadron fragmentation and decay chain is obtained.
The method is applied to $ep \rightarrow b\bar{b}X \rightarrow D^* \mu X$ and $\mu^+ \mu^-X$ final states at HERA. 
Comparisons of the data and NLO cross sections at visible and $b$-quark level were found to be consistent. \vspace{1.cm} \\
]

%%%%%%%%%%%%%%%%%%%%%%%%%%%%%%%%%%%%%%%%%%%%%%%%%%%%
\section{\textsf{Introduction}}
%%%%%%%%%%%%%%%%%%%%%%%%%%%%%%%%%%%%%%%%%%%%%%%%%%%% 
\label{sec-intro}
Heavy flavor (beauty and charm) production in $ep$ collisions at HERA allows QCD to be probed and understood in detail. 
Several heavy flavor production channels have been studied by the H1~\cite{H1} and ZEUS~\cite{ZEUS} experiments and compared to NLO 
QCD predictions based on the FMNR~\cite{FMNR}  
massive fixed order calculations. These calculations are currently 
the only ones available at NLO which include a fully differential description of the parton-level final states, and therefore a non-trivial propagation of cuts at the visible level (i.e.~on variables measured directly in the detector). In order to achieve this, the partons need to be 
fragmented and decayed to measurable hadron or lepton final states. For relatively simple cases, this can be implemented through a numerical or analytical parametrization of the parton-hadron level relation, obtained from separate MC simulations or direct measurements, which are then directly appended to the FMNR prediction.  
For final states with correlated cuts on several final particles, such as in~\cite{Chekanov:2006sg,Bloch:2005sy}, this is not easily possible,
although it has been attempted~\cite{Aktas:2005bt}. Nevertheless, NLO predictions for such final states are needed.\\

One solution to this problem is the MC@NLO approach~\cite{Frixione:2006he} --\ already implemented for the LHC and
 Tevatron\ -- which is not yet available for HERA. In this approach NLO QCD predictions 
are matched to parton showers in heavy flavor production, combining a Monte Carlo event
 generator (HERWIG~\cite{Corcella:1999qn}) with NLO calculations of rates for QCD processes.\\

The alternative proposed here is an interface of the FMNR program to P{\sc ythia}~\cite{Sjostrand:2001yu}.  The application of the REDSTAT option~\cite{nuncio} to the NLO description at quark level provided by FMNR, transforms FMNR into an effective Monte Carlo-like event generator. 
The parton-level events are then read back through the ``Les Houches accord''~\cite{hep-ph/0109068} interface into P{\sc ythia} 6.2, where the description of the hadron fragmentation and full decay chain is implemented.\\

No attempt is made to include additional parton showering. This will be achieved later by MC@NLO for HERA, when it will become available~\cite{HERAMCNLO}. 

%%%%%%%%%%%%%%%%%%%%%%%%%%%%%%%%%%%%%%%%%%%%%%%%%%%%%%%%%%%%%%%%%%%%%
\section{\textsf{The FMNR program}}
\begin{figure}[ht]
\resizebox{0.20\textwidth}{!}{%
  \includegraphics{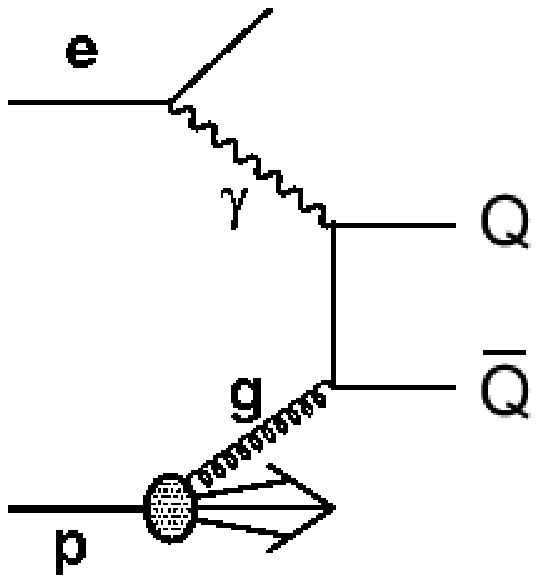}}
\hfill
\resizebox{0.23\textwidth}{!}{%
  \includegraphics{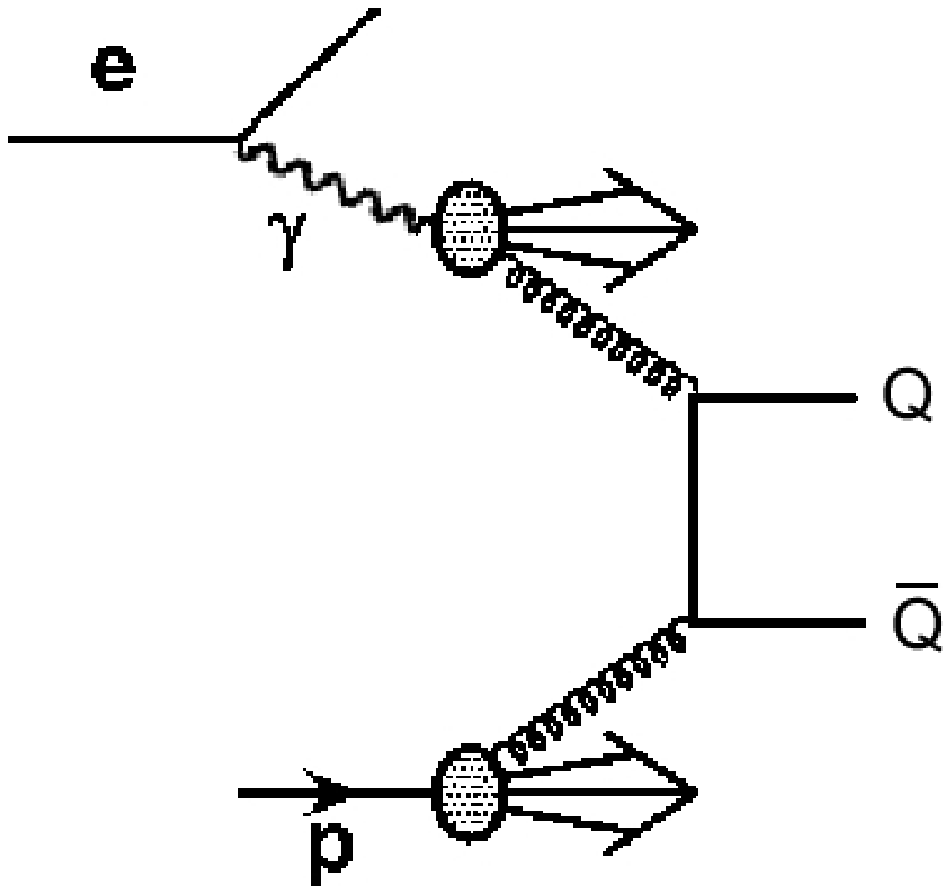}}
\caption{Point-like (left) and hadronic (right) photon coupling to the heavy quark in the leading order (LO) boson-gluon fusion process.}
\label{fig:point-hadronic}
\end{figure}

%%%%%%%%%%%%%%%%%%%%%%%%%%%%%%%%%%%%%%%%%%%%%%%%%%%%%%%%%%%%%%%%%%%%%%
\label{sec-fmnr}
FMNR makes cross-section calculations at NLO in QCD for heavy quark production in $ep$ and $\gamma p$ collisions. It is designed to give predictions in the photoproduction regime (photon virtuality $Q^2\lesssim 1$ GeV$^2$). It contains point-like and hadronic photon coupling to the heavy quarks (see Fig.~\ref{fig:point-hadronic}) in the massive fixed-order approach. In this scheme, the quarks are produced dynamically (i.e.~they are not included in the structure functions) and the relevant scale is the mass of the heavy quarks ($m_{\rm Q}, {\rm Q}=c,b$), neglecting  terms like $[ \alpha_s \ln{\mu ^2 / m^2_{\rm Q}}]^n$ that are resummed in other approaches. This makes this approach particularly valid for processes in which the hard scale $\mu^2 \sim m^2_{\rm Q}$. \\
%-------------------------------------------------------------------
%-------------------------------------------------------------------
 
In addition to quark-level predictions, the FMNR program provides a framework to fragment e.g.~$b$ quarks into $B$ hadrons, and simulate the decay of these hadrons by interfacing them to appropriately chosen decay spectra. However, decays to complex final states, like $D^* + \mu$ from the decay of the same $B$ hadron with cuts on both particles, cannot be easily implemented in this scheme.\\

\begin{figure}[htp]
\begin{center}
\resizebox{0.45\textwidth}{!}{%
  \includegraphics{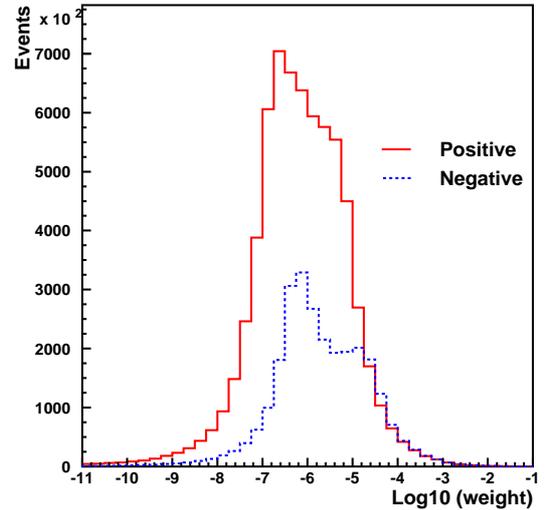}}
\end{center}
\caption{An example of distributions of positive (continuous line) and negative (dashed line) weights for events generated using FMNR.}
\label{fig:weights}
\end{figure}

A straightforward interface of the parton-level events produced by FMNR to a Monte Carlo-like fragmentation and simulation chain is not practical since weights (either positive or negative) are assigned to the output events. These weights range over more than 8 orders of magnitude as is shown in Figure~\ref{fig:weights}.  FMNR also differs from a true Monte Carlo generator because the events are not created randomly, but rather in a systematic order, and thus the complete data set must be considered in order to get meaningful results.  This  makes such an approach extremely inefficient because high statistics needs to be generated in order to keep statistical fluctuations low.\\
%--------------------------------------
%-------------------------------------- 
% deja un renglon libre aqui !

\begin{figure}[htp]
\begin{center}
\resizebox{0.45\textwidth}{!}{%
  \includegraphics{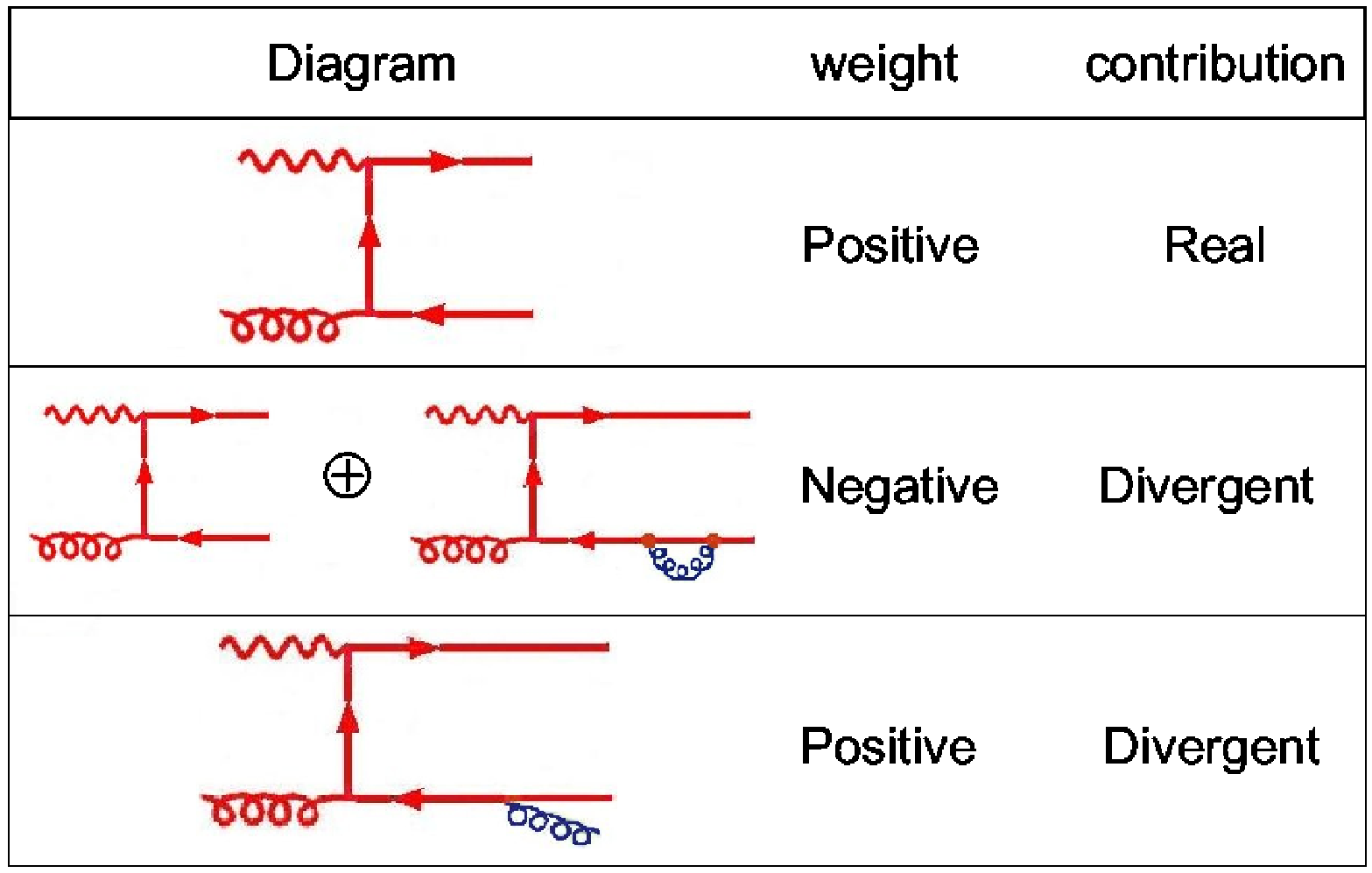}}
\end{center}
\caption{Examples of processes with positive or negative weight, and their corresponding contribution to the cross-section in the FMNR calculations.}
\label{fig:diagrams}
\end{figure}

As seen in Figure~\ref{fig:weights}, in most cases the weights assigned to the events are within four orders of magnitude. However, there are also cases in which a pair of similar events are created with relatively large (but oppositely signed) weights. This is due to the fact that in the phase-space region where FMNR performs the calculations, collinear and infrared  divergences are present \cite{Mangano:1991jk}.  Physically, these paired events correspond to two distinct but similar processes: In the first case, a quark-antiquark pair and a gluon are produced in the collision, and the gluon is such that it remains near one of the heavy quarks. In the second case, the gluon not only remains near the heavy quark but also at some later time is reabsorbed (see Figure ~\ref{fig:diagrams}).\\
%--------------------------------------
%-------------------------------------- 

For the cross-section calculation this means in practice that the process with e.g.~the collinear gluon has positive weight, but gives a divergent contribution, and the interference term between the LO process and the process with a virtual gluon has a negative weight and gives a divergent contribution as well. 
FMNR overcome the problem of the soft and collinear divergences by generating sequences of topologically correlated events in such a way that the cancellation is performed by the integration over that phase-space region. The output of FMNR is hence, weighted partonic events with a heavy quark-antiquark pair, and events with the pair plus an extra parton (a gluon or a light quark or anti-quark).\\

 However, problems do occasionally arise because the histograms used to plot the output are discretely binned, and it can happen that one of the two paired events will end up in one histogram bin, while the other event with only very slightly different topology ends up in the adjacent one. When this happens the two events can not cancel one another, and so the two adjacent bins are left with relatively large weighted events. It is then necessary to produce many more events with smaller weights for the histogram to smooth out.  Although the contributions to the cross section are finite at this point, the values of the weights for the generated events span over many orders of magnitude (Fig.~\ref{fig:weights}).\\  

A solution to this problem  is to find all paired events and suitably average them before they are output into the histogram. This is what the reduced statistics (REDSTAT) option does, as will be explained in detail in the following section.

%%%%%%%%%%%%%%%%%%%%%%%%%%%%%%%%%%%%%%%%%%%%%%%%%%%%%%%%%%%%%%%%%%%%%%%%%%%%%%
\section{\textsf{REDSTAT}}
%%%%%%%%%%%%%%%%%%%%%%%%%%%%%%%%%%%%%%%%%%%%%%%%%%%%%%%%%%%%%%%%%%%%%%%%%%%%%
\label{sec:redstat}
The group of subroutines REDSTAT is implemented as an extension to the FMNR program.  It is designed to:
\begin{itemize}
\item reduce the range of weights for the generated events, such that an interface to fragmentation and simulation packages like P{\sc ythia} becomes realistic;
\item reduce the necessary statistics without losing NLO accuracy;
\item improve run-time efficiency.
\end{itemize}
The method is as follows: REDSTAT monitors all the parton level events as they are generated by FMNR, and searches for events with weights above a given threshold. Two or more events with
large but opposite sign weights and similar kinematics are then combined into a new single event by averaging the 
four-momentum of the partons and assigning the sum of the weights as the new weight.
For events with weights below the threshold, REDSTAT makes a random decision to keep the event with a probability 
proportional to its weight (sampling approach). These events then acquire the weight of the threshold. \\

The threshold is chosen such that the number of events with positive weights is almost equal to the number of events with negative weights above that threshold. 
Events are considered to have similar kinematics when the difference in transverse momentum ($p_T$), rapidity ($\zeta$) and azimuthal angle ($\phi$) of the quark-antiquark pair are less than user cut values that should reflect the experimental detector resolution. \\

\begin{figure}[htp]
\begin{center}
\resizebox{0.45\textwidth}{!}{%
  \includegraphics{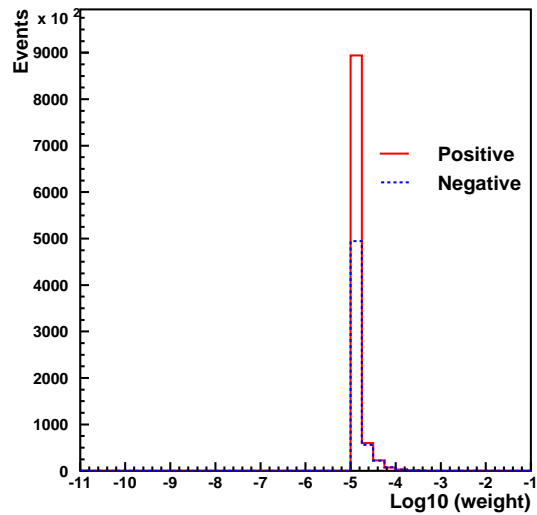}}
\end{center}
\caption{Distribution of weights for the generated FMNR events after REDSTAT.}
\label{fig:redstat}
\end{figure}

After this procedure, as can be seen in Figure~\ref{fig:redstat}, the range of weights is reduced to about two orders of magnitude, and the number of generated events is also reduced, therefore improving the run-time efficiency of subsequent applications using these events as input. \\

%--------------------------------------
%-------------------------------------- 

This method preserves the NLO accuracy for the relevant spectra (e.g. $b$-quark $p_T$ and angular distributions) and cross sections at parton level, as long as the chosen binning is not smaller than the assumed resolution, as will be explained in the next section. \\

Some caveats of this method are related to technical details of the REDSTAT implementation: the optimal threshold used to decide whether events will or will not be combined does not have a universal value. It must be chosen looking at the weight distributions, and these change depending on the generated process (charm or beauty) and according to the number of generated events (for example, it moves to the left when more statistics is generated). The default cut-off values of $p_T$, $\zeta$ and $\phi$ are tuned to the ZEUS detector resolution and have to be appropriately retuned for other applications.  Therefore, we recommend to contact the authors for advice before using REDSTAT.\\

Finally, an optional output file for the interface to other packages is created.  This is an ASCII file containing the 
information of the generated parton level events.  For each event, this includes e.g.~the assigned weight and the four-momenta 
of the two or three final state partons.

%%%%%%%%%%%%%%%%%%%%%%%%%%%%%%%%%%%%%%%%%%%%%%%%%%%%%%%%%%%%%%%%%%%%%%%%%%%%%%
\subsection{\textsf{Examples of REDSTAT results}}
%%%%%%%%%%%%%%%%%%%%%%%%%%%%%%%%%%%%%%%%%%%%%%%%%%%%%%%%%%%%%%%%%%%%%%%%%%%%%%
\begin{itemize}
\item {\bf Cross-section prediction at $\mathbf{b}$-quark level for the process $\mathbf{\gamma p \rightarrow b (\bar{b}) X}$ }\\

This was calculated using the original FMNR program and compared to the FMNR version which includes REDSTAT.  The kinematic region was $Q^2 < 1 $ GeV$^2$ and inelasticity $0.05 < y < 0.85$. \\

The parameters used in both calculations were:
  \begin{itemize}
  \item the mass of the $b$-quark $m_b = 4.75$ GeV;
  \item the renormalization and factorization scales, defined as $\mu = \sqrt{m^2_b + p^2_{Tb}}$, where $p_{Tb}^2$ is the average of the squared transverse momentum of the two emerging $b$ quarks;
  \item the structure functions CTEQ5M \cite{Lai:1999wy} for the proton and GRV-G-HO \cite{Gluck:1991ee} for the photon.
  \end{itemize}
For the evaluation of the theoretical uncertainty (relevant for section 
\ref{sect:examples}), the scales were varied by a 
factor 2, and the mass of the beauty quark was varied between 4.5 and 5 GeV. The variations were
done simultaneously such that the spread was maximized. The uncertainty of the
structure functions turned out to be small in comparison, and is neglected.\\

Additional REDSTAT parameters were:
  \begin{itemize}
  \item weight threshold: $10^{-5}$;
  \item maximal difference in transverse momentum between the partons of the events to be combined, $\Delta p_T = 1.0$ GeV;
  \item similarly, maximal difference in rapidity: $\Delta\zeta = 0.20$;
  \item maximal difference in the angle between the quarks: $\Delta\phi = 0.3$ rad.
  \end{itemize}
These cuts were applied to the two heavy (anti)quarks only.
The number of events generated in each sample are displayed in Table~\ref{tab:samples}.\\
%==================================
\begin{table}[htp]
\begin{center}
\begin{tabular}{lr}
\hline\noalign{\smallskip}
Sample                & Number of events  \\[1ex]
\hline\noalign{\smallskip}
FMNR (original)      &  31318299 \\[1ex]
FMNR with REDSTAT    &   3159440 \\[1ex]
FMNR high statistics & 322391999 \\[1ex]
FMNR fast REDSTAT    &   3077217 \\[1ex]
\hline\noalign{\smallskip}
\end{tabular}
\end{center}
\caption{Number of events generated per sample.}\label{tab:samples} 
\end{table}
%==================================
 
The results shown in Table~\ref{tab:xsections} demonstrate that the NLO accuracy of the original FMNR is preserved after the application of the REDSTAT option. \\

%==================================
\begin{table}[htp]
\begin{center}
\begin{tabular}{lc}
\hline\noalign{\smallskip}
{}                 & $\sigma_{\gamma p \rightarrow b (\bar{b}) X}$  \\[1ex]
\hline\noalign{\smallskip}
FMNR (original)    & 4.95 nb \\[1ex]
FMNR with REDSTAT  & 4.94 nb \\[1ex]
\hline\noalign{\smallskip}
\end{tabular}
\end{center}
\caption{Comparison of NLO cross-section predictions using FMNR without/with the REDSTAT option, for rapidity of the $b$-quark or antiquark  $|\zeta| < 1$.}\label{tab:xsections} 
\end{table}
%==================================

\item {\bf Comparison of the transverse momentum distribution of the $\mathbf{b(\bar{b})}$-quark}\\ 

As it is shown in Figure~\ref{fig:pt_comparison}, the REDSTAT option yields a good description of the $b$-quark $p_T$. The small differences in the shapes of the distributions are due to statistical fluctuations of such a typical FMNR run.\\

This is illustrated further in the Figure~\ref{fig:triple-pt-comparison}; in which in order to estimate the statistical fluctuations, 10 samples of standard FMNR runs where generated using the same statistics changing only the seed of the random number generator. Obtaining a distribution as the mean value of the 10 samples, the error of which has a variance $V(\bar x) = \sigma^2 / N$. This mean distribution (called standard in the Figure) is compared to a run using 10 times more statistics and to the version using REDSTAT. 
After REDSTAT, the fluctuations are expected to be smaller than in the 
original FMNR, since the probability to split a
correlated pair of events at the bin boundary is reduced. 
%--------------------------------------
\begin{figure}[htbp]
\begin{center}
\resizebox{0.45\textwidth}{!}{%
  \includegraphics{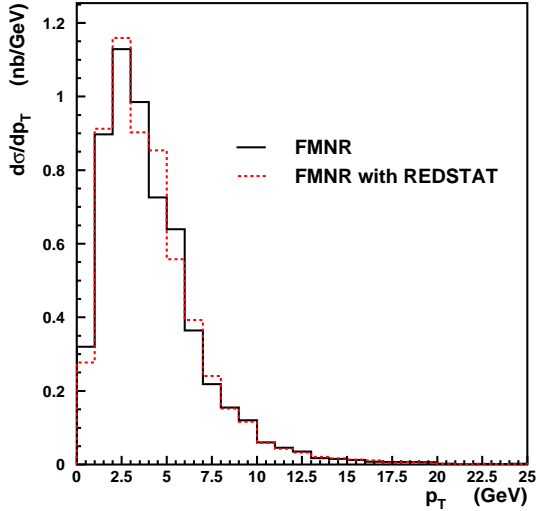}}
\end{center}
\caption{Comparison of the $b(\bar{b})$-quark transverse momentum distributions using the original FMNR (continuous line) and FMNR with REDSTAT (dashed line).}
\label{fig:pt_comparison}
\end{figure}
%-------------------------------------- 
 %--------------------------------------
\begin{figure}[htbp]
\begin{center}
\resizebox{0.45\textwidth}{!}{%
  \includegraphics{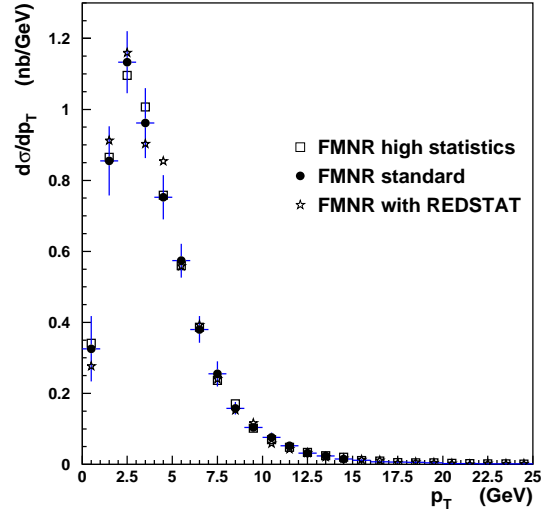}}
\end{center}
\caption{$b(\bar{b})$ transverse momentum distributions of the original FMNR with 10 times more statistics (squares), a mean of 10 FMNR samples (full circles) where the error bars are the statistical errors properly accounted for the different event weights, and FMNR using REDSTAT (stars).}
\label{fig:triple-pt-comparison}
\end{figure}
%-------------------------------------- 

\item {\bf Variation of detector resolution parameters}\\

A comparison between a run using default values tuned for ZEUS ($\Delta p_T = 1.0 $ GeV, $\Delta \zeta = 0.2$ and $\Delta \phi = 0.3$ rad) and a run with wider cut values ($\Delta p_T = 1.5$ GeV, $\Delta \zeta = 0.38$ and $\Delta \phi = 0.45$ rad) which therefore allows a faster performance, is shown in Figure~\ref{fig:detector-resolution}. The fluctuations result from the differences in the event combination procedure, since the same original FMNR events are generated in each case.

%--------------------------------------
\begin{figure}[htp]
\begin{center}
\resizebox{0.45\textwidth}{!}{%
   \includegraphics{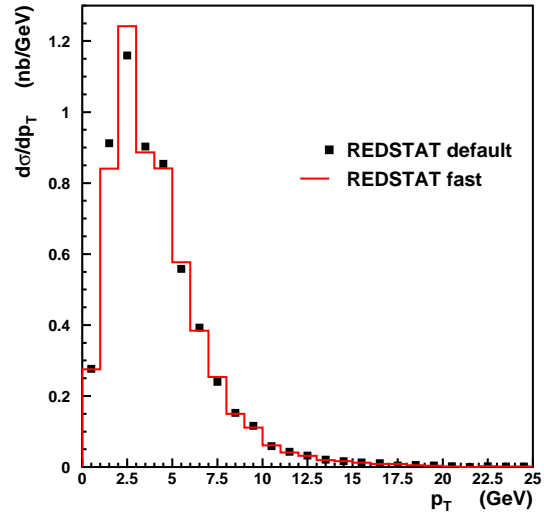}}
\end{center}
\caption{Comparison of the $b(\bar{b})$-quark transverse momentum distributions using REDSTAT with default parameters [1 GeV, 0.2, 0.3 rad] and REDSTAT with wider cut values [1.5 GeV, 0.38, 0.45] (fast).}
\label{fig:detector-resolution}
\end{figure}
%-------------------------------------- 
\end{itemize}

%%%%%%%%%%%%%%%%%%%%%%%%%%%%%%%%%%%%%%%%%%%%%%%%%%%%%%%%%%%%%%%%%%%%%%%%%%%%%%
\section{\textsf{The FMNR$\otimes$P{\sc ythia} Interface}}
%%%%%%%%%%%%%%%%%%%%%%%%%%%%%%%%%%%%%%%%%%%%%%%%%%%%%%%%%%%%%%%%%%%%%%%%%%%%%%
The implementation of the interface of the FMNR parton-level predictions to the fragmentation and decay chain from 
P{\sc ythia}/J{\sc etset} is a two-step process.
The first step of this \fxp interface consists in the application of the REDSTAT extension to the FMNR program (described in Section~\ref{sec:redstat}).  REDSTAT is used to transform FMNR into an effective Monte Carlo-like parton-level event generator, through the combination of events with similar kinematics and a sampling approach.\\

 The events obtained this way are written to an output file.
In the second step, these events are read back into the P{\sc ythia} 6.2 program through the ``Les Houches accord'' user interface.\\

The interface routine UPEVNT then sequentially reads one event from the input ASCII file and fills it into 
      the ``Les Houches'' interface variables to be used as input by P{\sc ythia}.
      At this stage, a ``reasonable'' (i.e.~physically possible) colour flow 
      is assigned to each FMNR parton level process. FMNR does not provide 
      this information, which is needed in the case of string fragmentation.
      The difference due to different possible colour flow assignments should 
      be included in the systematic error if this is critical for the application.\\

Since the first comparisons with data were made to measured beauty production cross sections \cite{}, the following sections are written for the specific case of beauty production. All arguments hold in an analog way for charm production.\\
  
The initial state partons are allowed to have an intrinsic $k_T$ (typically $\sim$ 300 MeV) as implemented in P{\sc ythia}.  This has a negligible effect on the resulting cross sections ($\sim 1\%$).\\

 Parton showering is {\em not} allowed in order to avoid double counting of higher order contributions. \\

Fragmentation of $b$-quarks close to production threshold turned out to be a non trivial issue. Since the details of the threshold treatment were found to be much more important than the choice of a particular fragmentation function, the Peterson formula with $\epsilon = 0.0035$ is used for convenience. Three approaches have been considered:

\begin{itemize}
\item{Independent fragmentation in the P{\sc ythia} model, within \fxpnospace.\\ 
This is used because FMNR does not provide color connections on an event-to-event basis, and color connections are not required in this model.}
\item{Fragmentation in the Lund string model, within \fxpnospace.\\  
For this, reasonable color connections have to be associated to each FMNR event.}
\item{Independent fragmentation scheme as provided in the context of the 
original FMNR.\\
This implies setting the $B$-hadron momentum equal to the $b$-quark momentum before reducing it according to the Peterson formula.}
 This neglects threshold corrections due to the need of simultaneous conservation of energy and momentum. 
\end{itemize}
In the first two cases, the fragmentation, decay tables and kinematics as implemented in P{\sc ythia} 6.2/J{\sc etset} are used to obtain a hadron-level event. Therefore, non-dominant arbitrarily complicated decays, such as $B \rightarrow D^* D$ followed by a $D \rightarrow \mu X$, or muons through intermediate $J/ \Psi$ or $\tau$ states, are automatically included.  The branching ratios were empirically corrected at analysis level to correspond to those obtained from the Particle Data Group (PDG) \cite{Yao:2006px}.
Unless otherwise stated, the 2$^{nd}$ approach is used for the central 
prediction, while the first enters the systematic uncertainty. \\

The third case could be used only for relatively simple final states 
(e.g. independent production of a muon from each heavy quark), and the 
resulting differences with respect to the first two cases were extrapolated 
to the more complicated ones (e.g. correlated production of two muons from 
the same $b$ quark), and used as a systematic check. 

%%%%%%%%%%%%%%%%%%%%%%%%%%%%%%%%%%%%%%%%%%%%%%%%%%%%%%%%%%%%%%%%%%%%%%%%%%%%%%
\section{\textsf{Examples of the use of the \fxp interface.}}\label{sect:examples}
%%%%%%%%%%%%%%%%%%%%%%%%%%%%%%%%%%%%%%%%%%%%%%%%%%%%%%%%%%%%%%%%%%%%%%%%%%%%%%
\label{sec:2examples}
Double tagging techniques like $\mu^+ \mu^-$ or $D^* \mu$ \cite{Chekanov:2006sg,Bloch:2005sy,Aktas:2005bt} benefit from the increasing amount of data 
provided by HERA. 
At the ZEUS experiment, a beauty enriched data sample is obtained in these channels taking advantage of the significantly reduced background. This allows softer kinematic cuts which, 
in combination with the wide rapidity coverage of the ZEUS muon detectors, results
 in an enhanced sensitivity to {\it B} hadrons produced at low
 transverse momenta, where the bulk of the $b$ cross section is concentrated.
Under these conditions a direct measurement of the total $b \bar{b}$ cross section
 becomes possible without the use of large model-dependent extrapolation factors.\\

Tagging muons or $D^*$ from different {\it b} quarks allows $b \bar{b}$ correlations to be explicitly measured, while tagging two muons or a muon and $D^*$ from the same $b$ quark yields measurements which are 
almost insensitive to such correlations. 
However, the complexity of the decays of these channels and the different cuts applied to the final state particles makes 
a calculation of the NLO prediction using the original FMNR approach very difficult.\\

In this section our new method will be applied to obtain NLO beauty cross-section predictions for the final state channels 
$D^* \mu $ and $\mu^+ \mu^-$.

%%%%%%%%%%%%%%%%%%%%%%%%%%%%%%%%%%%%%%%%%%%%%%%%%%%%%%
\subsection{\textsf{The $\mathbf{ep \rightarrow b\bar{b}X \rightarrow D^* \mu X'}$ channel}}
%%%%%%%%%%%%%%%%%%%%%%%%%%%%%%%%%%%%%%%%%%%%%%%%%%%%%%
In the case of  $D^* + \mu$~\cite{Chekanov:2006sg} an unlike-sign combination of both particles is observed in the same detector hemisphere 
if they come from the same {\it b} quark, (mainly via $B^0 \rightarrow D^*\mu\nu_{\mu} $)
yielding a pure {\it b} sample. If they originate from different {\it b} quarks then the $\mu$ and $D^*$ are emitted into different hemispheres.
Both like and unlike sign charge combinations are allowed. 
In the unlike sign case only, the signature is similar to the one given by 
charm from different {\it c} quarks.
These differences can be used to extract the beauty signal. More details are given in \cite{Longhin:2003bx}.\\

The cross section for inclusive beauty production $ep \to b\bar b X \to D^*\mu X'$ in the kinematic range $p^{D^*}_{T}>1.9$ GeV, $-1.5<\eta^{D^*}< 1.5$, $ p^{\mu}_{T}>1.4$ GeV  and $-1.75<\eta^{\mu}<1.3$ was measured to be:
\begin{equation} 
\sigma_{vis} = 160 \pm 37(\mbox{stat.})^{+30} _{-57} (\mbox{syst.}) \; \mbox{ pb}. 
\label{eq:sigma_vis}
\end{equation}

%%%%%%%%%%%%%%%%%%%%%%%%%%%%%%%%%%%%%%%%%%%%%%%%%%%%%%%%
%\subsubsection{\textsf{Visible beauty cross-sections prediction for $\mathbf{ep \rightarrow eb\bar{b}X \rightarrow e D^* \mu X}$}}
%%%%%%%%%%%%%%%%%%%%%%%%%%%%%%%%%%%%%%%%%%%%%%%%%%%%%%%%%
%The measured visible cross-section Eq.~(\ref{eq:sigma_vis}) 
This is larger than, but still compatible with, the corresponding \fxp NLO prediction of 
\begin{equation} 
\sigma_{vis}^{NLO} =
67^{+20}_{-11} \rm{(NLO)}^{+ 13}_{- 9}\rm{(frag. \oplus br.)}~\rm{pb}.
\label{eq:sigma_vis_NLO}
\end{equation}
where the first error refers to the uncertainties of the FMNR parton
level calculation, and the second error refers to the uncertainties
related to fragmentation and decay.\\

Strictly speaking, the FMNR predictions are only valid for the photoproduction regime. Here, the Weizs\"acker-Williams (WW) approximation with an effective $Q^2 _{max} < 25 \gev^2$ cutoff~\cite{vonWeizsacker:1934sx,Williams:1934ad,Frixione:1993yw} was used to include the $\sim 15 \%$ DIS contribution for a combined cross section.
This NLO QCD prediction is also listed in Table~\ref{d*muxsections}. \\

%==========================================================
%   ZEUS table  2 columns
%==========================================================
\begin{table*}[htbp]
\caption{Comparison of measured and predicted cross sections. For the measured, the first error is statistical and the second systematic. The NLO predictions at visible level were obtained using \fxpnospace, and the measurement at {\it b}-quark level was obtained extrapolating with plain P{\sc ythia}.}
\label{d*muxsections}
\begin{center}
\begin{tabular}{ccllc}
\hline\noalign{\smallskip}
cross-section    &  & measured                   & NLO QCD               & ratio                \\
\hline\noalign{\smallskip}
Visible & inclusive  & $160 \pm 37^{+30} _{-57}$ pb     & $67^{+24} _{-14}$ pb  & $2.4^{+0.9} _{-1.3}$  \\
        & $\gamma p$ & $115 \pm 29^{+21} _{-27}$ pb     & $54^{+18} _{-12}$ pb  & $2.1^{+0.8} _{-1.0}$  \\
\hline\noalign{\smallskip}
b level & $\gamma p$ & $11.9 \pm 2.9^{+1.8} _{-3.3}$ nb & $5.8^{+2.1} _{-1.3}$ nb & $2.0^{+0.8} _{-1.1}$  \\
\hline\noalign{\smallskip}
\end{tabular}
\end{center}
\end{table*}
%==========================================================

A cross section for the same kinematic range, but adding a photoproduction requirement ($Q^2 < 1$ GeV$^2$, $0.05<y<0.085$)
was also obtained. The result, as well as the corresponding \fxp prediction, are shown in Table~\ref{d*muxsections}.
As in the inclusive case, the prediction underestimates the measured cross section, but is compatible with the measurement within the large errors. \\

Finally, these measured visible-level cross sections were extrapolated to {\it b}-quark level using 
P{\sc ythia}.  
The NLO QCD prediction can then be obtained at parton level directly from 
the original FMNR calculation. 
From the comparison of the ratios at visible and {\it b}-quark level in 
Table~\ref{d*muxsections}, one can conclude that the P{\sc ythia} 
extrapolation, which was the only way to compare the cross sections
before Eq.~(\ref{eq:sigma_vis_NLO}) could be calculated \cite{Longhin:2003bx},
was reliable.

%%%%%%%%%%%%%%%%%%%%%%%%%%%%%%%%%%%%%%%%%%%%%%%%%
\subsection{\textsf{Comparison ZEUS - H1}}
%%%%%%%%%%%%%%%%%%%%%%%%%%%%%%%%%%%%%%%%%%%%%%%%%
The H1 Collaboration has measured a cross section similar to the ZEUS photoproduction cross section, in a slightly different kinematic region~\cite{Aktas:2005bt}: 
$p^{D^*} _{T} > 1.5$ GeV, $-1.5 < \eta^{D^*} < 1.5$, $ p^{\mu} > 2.0$ GeV, $-1.735 < \eta^{D^*} < 1.735$, $Q^2 < 1$ GeV$^2$ and $0.05<y<0.75$. Its value is shown in Table~\ref{H1xsections}.
Using the \fxp interface, the ZEUS cross section can be extrapolated to the H1 kinematic region~\cite{Chekanov:2006sg}. A direct comparison of the two results is shown in Table~\ref{H1xsections}. Reasonable agreement is found.\\

%==========================================================
%   H1
%==========================================================
\begin{table}[ph]
\caption{Comparison of H1 and ZEUS visible cross section for $ep \rightarrow b\bar{b}X \rightarrow D^* \mu X'$. The ZEUS cross section has been extrapolated 
to the H1 kinematic range.} 
\label{H1xsections}
\begin{tabular}{rcc}
\hline\noalign{\smallskip}
cross-section & H1 & ZEUS (extrap.)\\
\hline\noalign{\smallskip}
$\gamma p$ & $206 \pm 53 \pm 35$ pb & $135 \pm 33^{+24} _{-31}$ pb  \\
\hline\noalign{\smallskip}
\end{tabular}
\end{table}
%==========================================================

The corresponding NLO prediction from \fxp is  
\begin{equation} 
\sigma_{vis,H1}^{NLO} =
61^{+17}_{-12} \rm{(NLO)}^{+ 12}_{- 8}\rm{(frag. \oplus br.)}~\rm{pb}.
\label{eq:sigma_vis_H1}
\end{equation}

The data to NLO ratio is again consistent with the ones quoted in  
Table~\ref{d*muxsections}.\\

The NLO prediction is somewhat larger than the one evaluated in \cite{Aktas:2005bt}
due to the inclusion of the hadron-like photon contribution, the inclusion of 
secondary-muon branching fractions for $D^*$ and $\mu$ from the same $b$ quark
(e.g. $B \to D^* D \to D^*\mu X$),
which are difficult to handle outside the \fxp framework, and a detailed 
simulation of the kinematics of the $b \to B \to D^*$ chain rather than direct 
collinear fragmentation of $b$ quarks into $D^*$ mesons. 
This reduces the discrepancy claimed in \cite{Aktas:2005bt} and is an example 
for the importance of a detailed treatment of complicated final states.

%%%%%%%%%%%%%%%%%%%%%%%%%%%%%%%%%%%%%%%%%%%%%%%%%%%%%%%%%%%%%%%%%%%%%%%%%%%
\subsection{\textsf{The $\mathbf{ep \rightarrow b\bar{b}X \rightarrow \mu^+ \mu^- X'}$ channel}}
%%%%%%%%%%%%%%%%%%%%%%%%%%%%%%%%%%%%%%%%%%%%%%%%%%%%%%%%%%%%%%%%%%%%%%%%%%%
 The signature of the dimuon channel is very similar to the $D^* \mu$ channel
% shown in Figure~\ref{fig:dstarmu-finalstates}, 
and detailed information can be found in~\cite{Bloch:2005sy}.
Again, this is a very complicated final state, for which the cross section
measurements are being finalized by ZEUS.
The purpose of this section is to provide cross section predictions to which
the ZEUS results can be compared.
 
%%%%%%%%%%%%%%%%%%%%%%%%%%%%%%%%%%%%%%%%%%%%%%%%%%%%%%%%%%%
\subsubsection{\textsf{Visible beauty cross-sections prediction for  $\mathbf{ep \rightarrow b\bar{b}X \rightarrow \mu^+ \mu^- X'}$}}
%%%%%%%%%%%%%%%%%%%%%%%%%%%%%%%%%%%%%%%%%%%%%%%%%%%%%%%%%%%%
The measured visible cross-section compared to the NLO prediction from \fxp is shown in Table~\ref{dimuxsections}. The extrapolation to {\it b}-quark level was done using P{\sc ythia} and is compared to the NLO prediction, obtained by adding up the (original) FMNR and HVQDIS~\cite{Harris}  predictions, for the photoproduction and DIS regions respectively.\\

Here, as in the case of the $D^* \mu$ channel, the cross-section comparisons at visible and {\it b}-quark level are consistent.
%==========================================================
% dimu
%==========================================================
\begin{table*}
\caption{Comparison of measured and predicted dimuon cross-sections. For the measured, the first error is statistical and the second systematic.}
\label{dimuxsections} 
\begin{center}
\begin{tabular}{crrrc}
\hline\noalign{\smallskip}
{} & cross-section & measured (prel.) & NLO QCD & ratio \\
\hline\noalign{\smallskip}
Visible & Total           & $63 \pm 7^{+20} _{-18}$ pb       & $30^{+9} _{-6}$ pb      & $2.1^{+0.8} _{-1.0}$  \\
\hline\noalign{\smallskip}
$b$ level & Total           & $16.1 \pm 1.8^{+5.3} _{-4.7}$ nb & $6.8^{+3.0} _{-1.7}$ nb & $2.3^{+1.0} _{-1.2}$  \\
\hline\noalign{\smallskip}
\end{tabular}
\end{center}
\end{table*}
%==========================================================
%%%%%%%%%%%%%%%%%%%%%%%%%%%%%%%%%%%%%%%%%%%%%%%%%%%%%%%%%%%
\subsubsection{\textsf{Differential cross sections}}
\begin{figure}[htbp]
  \resizebox{0.45\textwidth}{!}{%
   \includegraphics{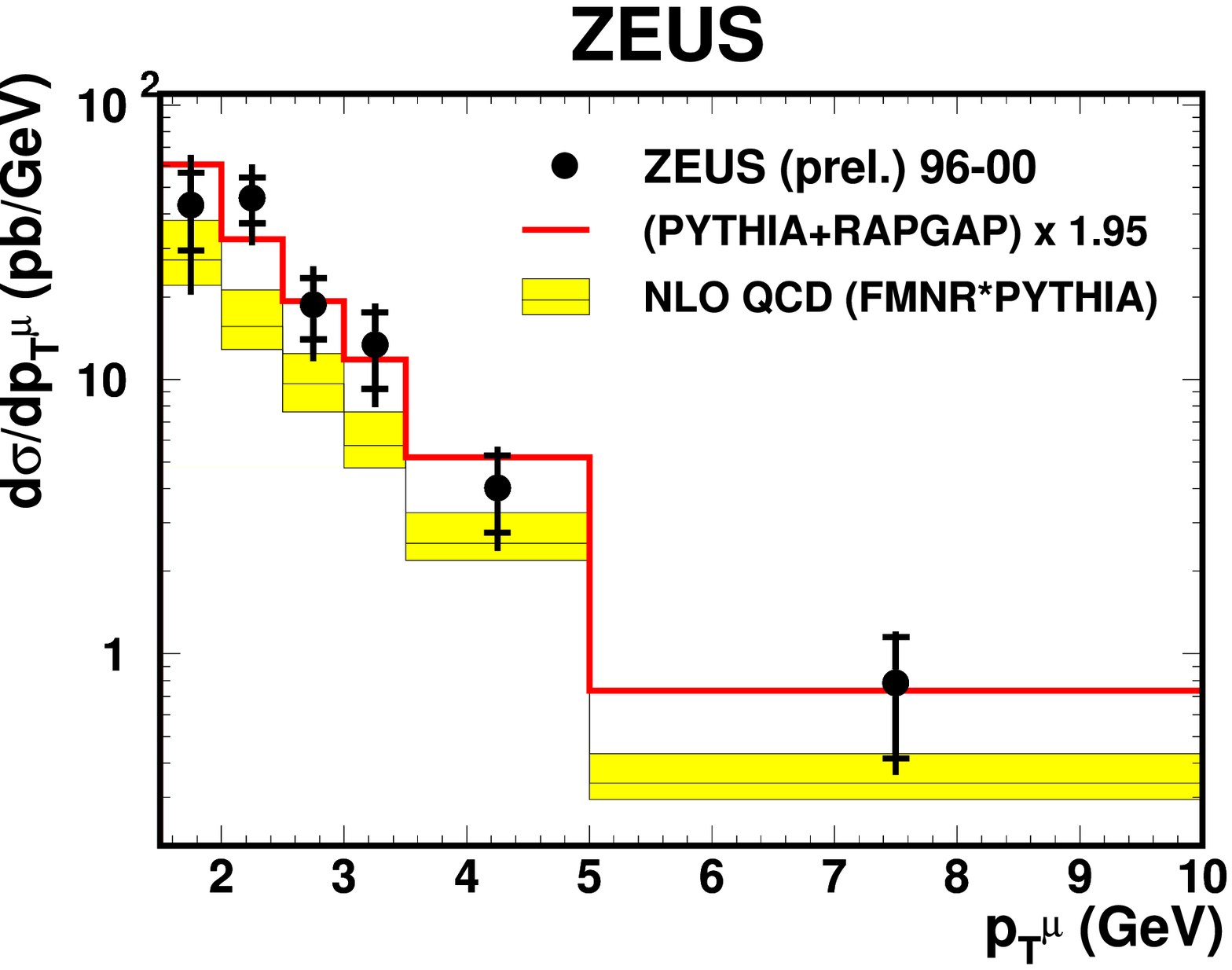}}
 \resizebox{0.45\textwidth}{!}{%
   \includegraphics{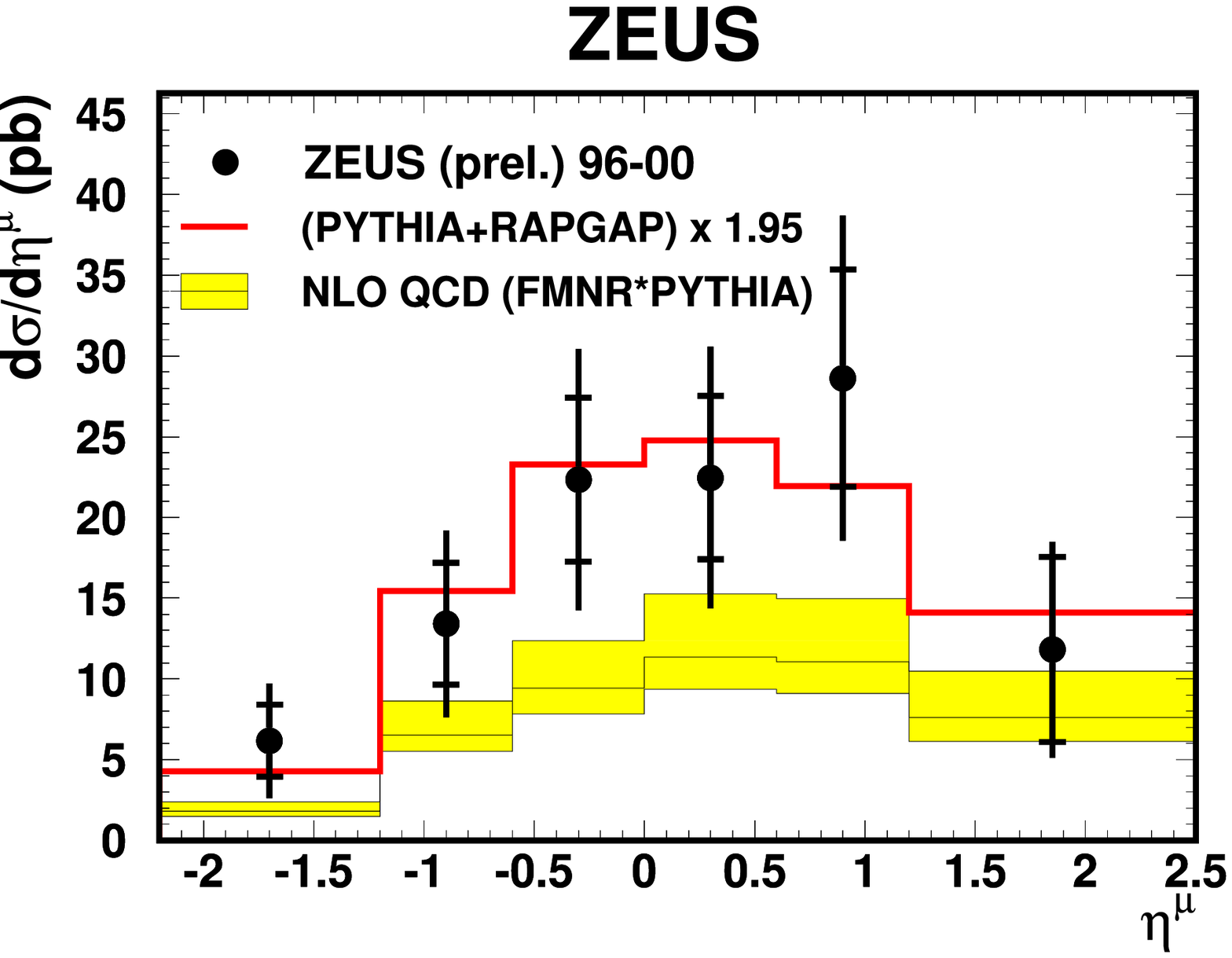}}
\caption{Differential cross-sections $d\sigma / dp_T$ (left) and  $d\sigma / d\eta$ (right) for muons. The data points are compared to the scaled LO prediction from P{\sc ythia} + R{\sc apgap}, and to the NLO prediction from \fxpnospace.}
\label{fig:pt-eta}
\end{figure}

%%%%%%%%%%%%%%%%%%%%%%%%%%%%%%%%%%%%%%%%%%%%%%%%%%%%%%%%%%%%
Differential cross-sections were also obtained and shown in Figure~\ref{fig:pt-eta}.  One can observe the general trend of NLO to lie below the data, but consistent between errors. The shape of the distributions is well reproduced by the NLO prediction. 
%-------------------------------------------------------------------
%-------------------------------------------------------------------

\begin{figure}[htbp]
%\begin{center}
\resizebox{0.45\textwidth}{!}{%
  \includegraphics{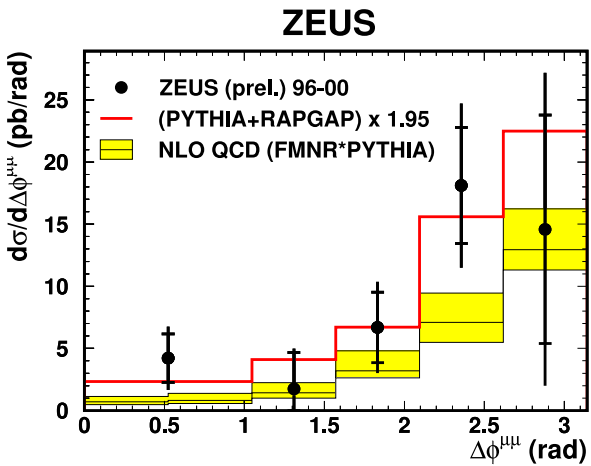}}
%\end{center}
\caption{Differential cross-section $d\sigma / \Delta \phi$ for muons from different {\it b}-quarks}
\label{fig:dphi}
\end{figure}

For the correlations between the two $b$-quarks, the reconstructed dimuon mass range was restricted to $m^{\mu\mu}>3.25$ GeV. This additional cut reduces the probability that both quarks come from the same B-hadron. The corresponding differential cross-section is shown in Figure~\ref{fig:dphi}. The distribution is reasonably described by the \fxp NLO prediction within the large errors. The leading order contribution alone, which is also shown, reproduces the measured less well. Although the difference in shape is not dramatic, this could confirm the importance of the contributions from higher order processes.   
%-------------------------------------------------------------------
%-------------------------------------------------------------------

\subsection{Other applications}
\label{sect:other}

In principle, the \fxp interface can be used to generate events which can be 
propagated through the full simulation and reconstruction chain of any 
experiment measuring $ep$ or $\gamma p$ heavy flavor final states (if the 
chain allows the handling of negative weights).
This opens up additional applications, such as cross sections for hadron-level
heavy flavor jets, with hadron to parton-level corrections fully consistent
with the NLO approach. So far, LO+PS Monte Carlos needed to be used for 
this purpose. \\  

Examples for potential applications in the near future include
\begin{itemize}
\item evaluation of parton to hadron-level corrections for measurements of
   the charm fragmentation function at HERA \cite{Aktas:2004ka,Padhi:2004ug};
\item evaluation of NLO cross sections for measured 
  $\mu\mu$-jet-jet and $\mu e$-jet-jet final states from both charm and beauty\cite{hep-ex/0311015,Goettlich:2007ij}, for which no NLO predictions exist so far;
\item reevaluation and verification of NLO predictions for essentially all 
  earlier HERA heavy flavor results \cite{H1,ZEUS} at visible level.   
\end{itemize}
These applications will not obliviate the need for calculations using the 
MC@NLO approach, which will be able to handle final state parton showering
in addition. However, they have the virtue of being available immediately.

%%%%%%%%%%%%%%%%%%%%%%%%%%%%%%%%%%%%%%%%%%%%%%%%%%%%
\section{\textsf{Conclusions}}
%%%%%%%%%%%%%%%%%%%%%%%%%%%%%%%%%%%%%%%%%%%%%%%%%%%%
Cross sections for heavy quark production in $ep$ collisions at HERA are being measured with high precision. This includes channels with complex final states for which NLO QCD predictions are not easily obtained from simple extensions to 
parton level calculations.  The \fxp interface allows such predictions to be obtained for the photoproduction regime.\\

In the $ep \rightarrow b\bar{b}X \rightarrow D^* \mu X'$ channel, both at visible and quark level, the measured data at ZEUS exceed the \fxp NLO prediction, but are compatible within the errors.  It was also found that the H1 and ZEUS visible cross sections are consistent.\\

The almost constant data to NLO ratio obtained from the cross sections shows that the extrapolation from visible to quark level, or vice versa, is meaningful and reliable.\\

Inclusive and differential cross sections in the $ep \rightarrow b\bar{b}X \rightarrow \mu^+ \mu^- X'$ 
channel are evaluated, which can be compared to ZEUS measurements, currently being finalized.
Other potential applications to both charm and beauty measurements 
in $ep$ and $\gamma p$ interactions have been outlined.

%%%%%%%%%%%%%%%%%%%%%%%%%%%%%%%%%%%%%%%%%%%%%%%%%%%%
\section{\textsf{Acknowledgments}}
%%%%%%%%%%%%%%%%%%%%%%%%%%%%%%%%%%%%%%%%%%%%%%%%%%%%
The authors are very grateful to the authors of the FMNR program for providing the source code, to Massimo Corradi for providing a copy of this original code and for encouraging interest and very helpful discussions, to Johann Gagnon-Bartsch for starting the REDSTAT project, to Detlef Bartsch for very productive discussions and to Stefano Frixione for advise and very useful critical remarks.

%%%%%%%%%%%%%%%%%%%%%%%%%%%%%%%%%%%%%%%%%%%%%%%%%%%%
% References
%%%%%%%%%%%%%%%%%%%%%%%%%%%%%%%%%%%%%%%%%%%%%%%%%%%%
%

\vfill\eject

%------------------------------------------------------------------------------
%       Bibliography
%------------------------------------------------------------------------------
{
\def\bibname{\Large\bf References}
\def\refname{\Large\bf References}
{\raggedright

}
}
\vfill\eject

%------------------------------------------------------------------------------
%       Tables
%------------------------------------------------------------------------------
%\include{FxP-paper-tab}
%------------------------------------------------------------------------------
%       Figures
%------------------------------------------------------------------------------
%\include{FxP-paper-fig}
%
%       ... that's it
%
\end{document}